\documentclass[a4paper]{jpconf}
\usepackage{amsfonts,amsmath,amssymb}
\usepackage{graphicx}

\newcommand{\be}{\begin{equation}}
\newcommand{\bea}{\begin{eqnarray}}
\newcommand{\ee}{\end{equation}}
\newcommand{\eea}{\end{eqnarray}}
\newcommand{\sla}{\slash \hspace{-0.22cm}}

\def\s#1{{\scriptscriptstyle #1}}

\def\1eq#1{Eq.~(\ref{#1})}
\def\2eqs#1#2{Eqs.~(\ref{#1}) and~(\ref{#2})}
\def\3eqs#1#2#3{Eqs.~(\ref{#1}), (\ref{#2}) and~(\ref{#3})}
\def\4eqs#1#2#3#4{Eqs.~(\ref{#1}), (\ref{#2}), (\ref{#3}) and~(\ref{#4})}
\def\noeq#1{(\ref{#1})}

\def\fig#1{Fig.~\ref{#1}}

\def\diff#1{{\rm d}^{#1}}

\def\pslash{p\hspace{-0.18cm}\slash}

\def\G{\Gamma}

\def\gA{g^2 C_A}

\def\pslash{p\hspace{-0.18cm}\slash}
\def\qslash{q\hspace{-0.19cm}\slash}

\def\ps#1{p\hspace{-0.18cm}\slash_#1}
\def\tslash{t\hspace{-0.18cm}\slash}
\def\pslash{p\hspace{-0.18cm}\slash} 
\def\tsigma{\widetilde{\sigma}} 
\def\K#1{K_{#1}}
\def\K#1#2{K_{#1}^{\s{#2}}}
\def\oK#1#2{\overline{K}_#1^{\s{#2}}}
\def\Ga#1#2{\Gamma_{#1}^{\s{#2}}}

\def\spr{\!\cdot\!}

\begin{document}
\title{A new method for computing the quark-gluon vertex }

\author{A.~C. Aguilar}

\address{University of Campinas - UNICAMP, Institute of Physics ``Gleb Wataghin'',\\
CEP 13083-859 - Campinas, SP, Brazil}

\ead{aguilar@ifi.unicamp.br}

\begin{abstract}

In this talk we present a new method for determining the nonperturbative quark-gluon vertex, which constitutes a crucial ingredient for a variety of theoretical and phenomenological studies. This new method relies heavily on the exact all-order relation connecting the conventional quark-gluon vertex with the corresponding vertex of the background field method, which is Abelian-like. The longitudinal part of this latter quantity is fixed using the standard gauge technique, whereas the transverse is estimated with the help of the so-called transverse Ward identities. This method allows the approximate determination of the nonperturbative behavior of all twelve form factors comprising the quark-gluon vertex, for arbitrary values of the momenta. Numerical results are presented for the form factors in three special kinematical configurations (soft gluon and quark symmetric limit, zero quark momentum), and compared with the corresponding lattice data.

\end{abstract}

\section{Introduction}

\bigskip
\indent

One of the major challenges of nonperturbative QCD is to understand the mechanism  that drives chiral symmetry breaking and the associated generation of constituent quark masses. As is well known from a series of previous studies~\cite{Maris:2003vk,Roberts:1994dr,Fischer:2003rp,Aguilar:2010cn,Cloet:2013jya}, the quantity that is intimately related to the underlying dynamics of the chiral symmetry breaking is the \emph{quark-gluon vertex}. In addition, this vertex is of paramount importance in the formalism of the  Bethe-Salpeter equations (BSEs), which describes the formation of the bound states of the theory~\cite{Maris:1999nt,Bender:2002as,Bhagwat:2004hn,Holl:2004qn,Chang:2009zb,Williams:2014iea}.

From the point of view of the perturbative QCD, the quark-gluon vertex 
has been carefully scrutinized at the one-loop level~\cite{Davydychev:2000rt}, where results for general kinematic configurations and arbitrary gauges. Moreover, at two and three-loop order we have results for some specific kinematics and gauges~\cite{Chetyrkin:2000fd,Chetyrkin:2000dq}.

The main difficulty in dealing with the quark-gluon vertex lies in the fact
that one has to determine the behavior of twelve form factors 
(four ``longitudinal'' and the eight ``transverse''), which 
are functions of three independent  kinematic variables. 
For this reason, the available nonperturbative information on this quantity is rather limited. In particular,
there are only few results obtained from simulations on  
relatively small lattices~\cite{Skullerud:2002sk,Skullerud:2002ge,Skullerud:2003qu,Skullerud:2004gp,Lin:2005zd,Kizilersu:2006et}. 
In the context of the Schwinger-Dyson equations (SDEs) the situation is not that different. The behavior of the form factors is governed by a complex system of coupled integral equations, which can be solved only after considerable truncations and drastic simplifying assumptions 
~\cite{Bhagwat:2004kj,LlanesEstrada:2004jz,Williams:2014iea,Alkofer:2008tt,Matevosyan:2006bk,Aguilar:2013ac,Rojas:2013tza}. 

In this talk we will present a novel nonperturbative approach for
calculating the form factors of the quark-gluon vertex in the Landau gauge~\cite{Aguilar:2014lha}.
This task will be accomplished within the PT-BFM  scheme~\cite{Aguilar:2006gr,Binosi:2007pi,Binosi:2008qk},
which is obtained   from  the    combination   of   the   pinch   technique
(PT)~\cite{Cornwall:1981zr,Cornwall:1989gv,Pilaftsis:1996fh,Binosi:2002ft,Binosi:2003rr,Binosi:2009qm}
with the background field method (BFM)~\cite{Abbott:1980hw}. 

An intrinsic feature of the PT-BFM formalism is the natural separation of the gluonic field into a “quantum”  
and a “background”  part, thus 
increasing the number of possible Green's functions that one may consider. In particular, two types of the quark-gluon vertices make their appearance: (i) the conventional quark-gluon vertex (formed by a quantum gluon, quark and anti-quark fields), denoted by 
$\Gamma$; and the background quark-gluon vertex (formed by a background gluon, quark and anti-quark fields), denoted by $\widehat\Gamma$.
A crucial difference between these two vertices, lies in the fact that 
the conventional vertex satisfies the usual STIs, whereas the background vertex obeys  Abelian-like WIs. 
The conversion between quantum and background vertices is achieved through the so-called background-quantum identities (BQIs)~\cite{Grassi:1999tp,Binosi:2002ez}, 
which relate $\Gamma$ and $\widehat\Gamma$
through special auxiliary ghost Green's functions, namely $\Lambda_{\mu\nu}$, $K_{\mu}$ and its conjugated  $\overline{K}_{\mu}$.

The aim of this work is to express the conventional quark-gluon vertex as a deviation from the ``Abelian-like'' vertex  $\widehat\Gamma_{\mu}$. Specifically, our strategy will be the following: first we use the ``gauge technique'' 
inspired Ansatz~\cite{Salam:1963sa,Salam:1964zk,Delbourgo:1977jc,Delbourgo:1977hq}   for the longitudinal part of the Abelian-like $\widehat\Gamma_{\mu}$. Then, with the help of the so-called ``Transverse Ward Identities''(TWIs) ~\cite{Takahashi:1985yz,Kondo:1996xn,He:2000we,He:2006my,He:2007zza}, we will fix the transverse part of the $\widehat\Gamma_{\mu}$, neglecting the non-local terms present in the TWIs. The  combination of these two steps generates the so-called minimal Ansatz for $\widehat\Gamma_{\mu}$~\cite{Qin:2013mta}.
As a final step we use the BQIs to convert $\widehat\Gamma_{\mu}$ into $\Gamma$. This conversion requires 
the computation of the special auxiliary functions  
$\Lambda_{\mu\nu} $ and $K_{\mu}$; the behavior of$\Lambda_{\mu\nu}$ is well-known from previous studies~\cite{Aguilar:2009pp}, whereas $K_{\mu}$ and $\overline{K}_{\mu}$ were first computed at one-loop dressed approximation in Ref.~\cite{Aguilar:2014lha}, using as ingredient the gluon lattice propagator.

In order  to make contact  with the results of 
lattice simulations~\cite{Skullerud:2002sk,Skullerud:2002ge,Skullerud:2003qu,Skullerud:2004gp,Lin:2005zd,Kizilersu:2006et},
 we will present the  numerical evaluation of the relevant  form factors in three  special   kinematical  configurations namely (i) soft gluon, (ii) quark symmetric  limit and (iii)  zero quark  momentum. As we will see, while a general qualitative agreement with the available  
lattice data is found, quantitative discrepancies still remain.

\section{\label{vertex}  The two quark-gluon vertices in the  PT-BFM formalism}

\bigskip
\indent

\begin{figure}[!t]
\begin{center} 
\includegraphics[scale=0.675]{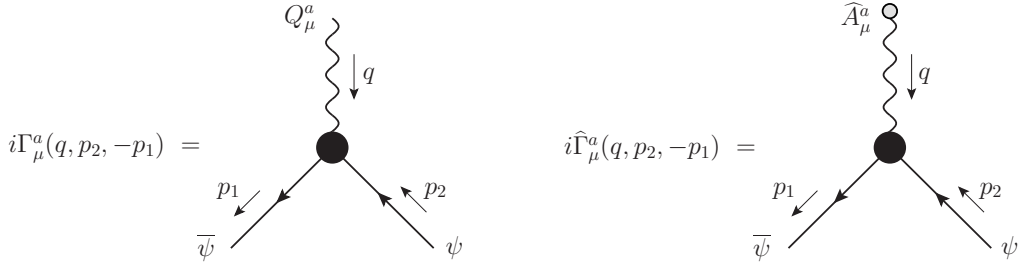}
\caption{\label{quark-vertex}The conventional and background quark-gluon vertex with the momenta routing used throughout the text.}  
\end{center}
\end{figure}

In the PT-BFM formalism there are two quark-gluon vertices, depending on the nature of the incoming gluon. Specifically, the vertex formed 
by a quantum gluon (Q) entering into a $ \psi \bar\psi$ pair corresponds to the conventional vertex, $\Gamma_\mu^a$ (see left vertex of~\fig{quark-vertex}); the corresponding three-point function with a 
background gluon ($\widehat{A}$) entering represents instead the PT-BFM vertex, and will be denoted by $\widehat{\Gamma}_\mu^a$ (see right vertex of~\fig{quark-vertex}).  
Choosing the flow of the momenta such that $p_1 = q+p_2$, we then define  
\be
i\Gamma_\mu^a(q,p_2,-p_1) = igt^a\Gamma_\mu(q,p_2,-p_1);\qquad 
i\widehat{\Gamma}_\mu^a(q,p_2,-p_1) = igt^a\widehat{\Gamma}_\mu(q,p_2,-p_1),
\ee
where the hermitian and traceless generators $t^{a}$ of the fundamental SU(3) representation are given by $t^{a} = \lambda^{a}/2$, with $\lambda^{a}$ the Gell-Mann  matrices.  

It is important to stress that \mbox{$\Gamma_\mu$ and $\widehat{\Gamma}_\mu$} coincide only at tree-level, where one has \mbox{$\Gamma^{(0)}_\mu = \widehat{\Gamma}^{(0)}_\mu = \gamma_\mu$}. 

\begin{figure}[!h]
\begin{center} 
\includegraphics[scale=0.5]{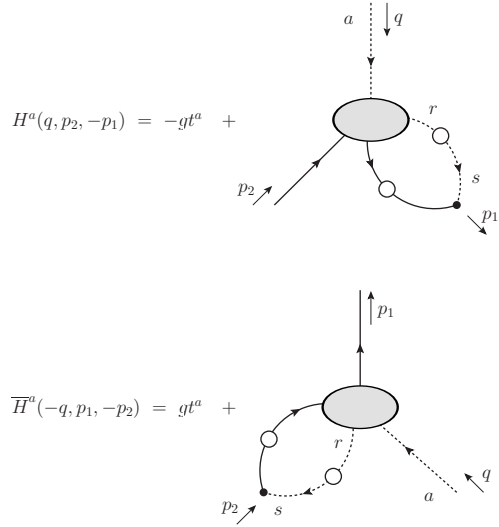}
\caption{\label{H-functions}The ghost kernels $H$ and $\overline{H}$ appearing in the STI satisfied by the quark vertex $\Gamma_\mu$. The composite operators $\psi c^s$ and $\bar\psi c^s$ have the tree-level expressions $-gt^a$ and $gt^a$ respectively.}  
\end{center}
\end{figure}

The essential difference between these two vertices is that $\widehat{\Gamma}_\mu$ obeys the QED-like WI~\cite{Abbott:1980hw}  
\be
q^\mu \widehat{\Gamma}_{\mu}(q,p_2,-p_1)= S^{-1}(p_1) - S^{-1}(p_2) , 
\label{WI}
\ee
instead of the standard STI
\be
q^\mu\Gamma_{\mu}(q,p_2,-p_1)=F(q^2)\left[S^{-1}(p_1)H(q,p_2,-p_1)-
\overline{H}(-q,p_1,-p_2)S^{-1}(p_2)\right],
\label{STI}
\ee
satisfied by $\Gamma_\mu$. In both expressions, \mbox{$S^{-1}(p)= A(p^2)\,\sla{p} - B(p^2),$} is  the inverse of the full quark propagator, defined in terms of the wave function, $A(p^2)$, and the mass function $B(p^2)$. In addition, 
$F(q^2)$ denotes the ghost dressing function, related to the full ghost propagator by $D(q^2)= F(q^2)/q^2$, whereas the functions $H^a=-gt^aH$ and its conjugated $\overline{H}^a=gt^a\overline{H}$ correspond to the so-called quark-ghost kernel, and are shown in~\fig{H-functions}. 

Notice that the quark-ghost kernel admits the general decomposition~\cite{Davydychev:2000rt}
\bea
H(q,p_2,-p_1)&=&X_0 \mathbb{I}+X_1 \ps{1}+X_2\ps{2}+X_3\tsigma_{\mu\nu}p^\mu_1p^\nu_2\,, \\ \nonumber
\overline{H}(-q,p_1,-p_2)&=&{\overline{X}}_0 \mathbb{I}+{\overline{X}}_2 \ps{1}
+{\overline{X}}_1\ps{2}+{\overline{X}}_3\tsigma_{\mu\nu}p^\mu_1p^\nu_2,
\label{HH}
\eea
where $\tsigma_{\mu\nu}=\frac{1}{2}[\gamma_\mu,\gamma_\nu]$ and  we have introduced the compact notation on the form factors \mbox{$X_i=X_i(q^2,p_2^2,p_1^2)$} and \mbox{$\overline{X_i}={X}_i(q^2,p_1^2,p_2^2)$}. Notice that at tree-level, one clearly has \mbox{$ X^{(0)}_0= \overline{X}^{(0)}_0=1$}, with the remaining form factors vanishing.

It is important to stress here that a set of identities, called background quantum identities (BQIs)~\cite{Grassi:1999tp,Binosi:2002ez}, 
relate the conventional and PT-BFM vertices. The BQI of interest in our case  
reads~\cite{Binosi:2008qk} 
\bea
\widehat{\Gamma}_\mu(q,p_2,-p_1)&=&\left[g^\nu_\mu \left(1+G(q^2)\right)+\frac{q_\mu q^\nu}{q^2}L(q^2)
\right]\Gamma_\nu(q,p_2,-p_1)\nonumber \\
&-&S^{-1}(p_1){K}_\mu(q,p_2,-p_1)-\overline{K}_\mu(-q,p_1,-p_2)S^{-1}(p_2),
\label{BQI}
\eea
where, in the Landau gauge,   the special functions ${K}_\mu$ and its conjugated $\overline{K}_\mu$ are related to the quark-ghost kernel $H$ (and its conjugated) in the following way   
\bea
H(q,p_2,-p_1) &=& 1 + q^\mu K_\mu(q,p_2,-p_1)\,, \nonumber   \\
\overline{H}(-q,p_1,-p_2)&=&  1 - q^\mu \overline{K}_\mu(-q,p_1,-p_2)\,.
\label{HK}
\eea

\begin{figure}[!h]
\begin{center} 
\includegraphics[scale=0.6]{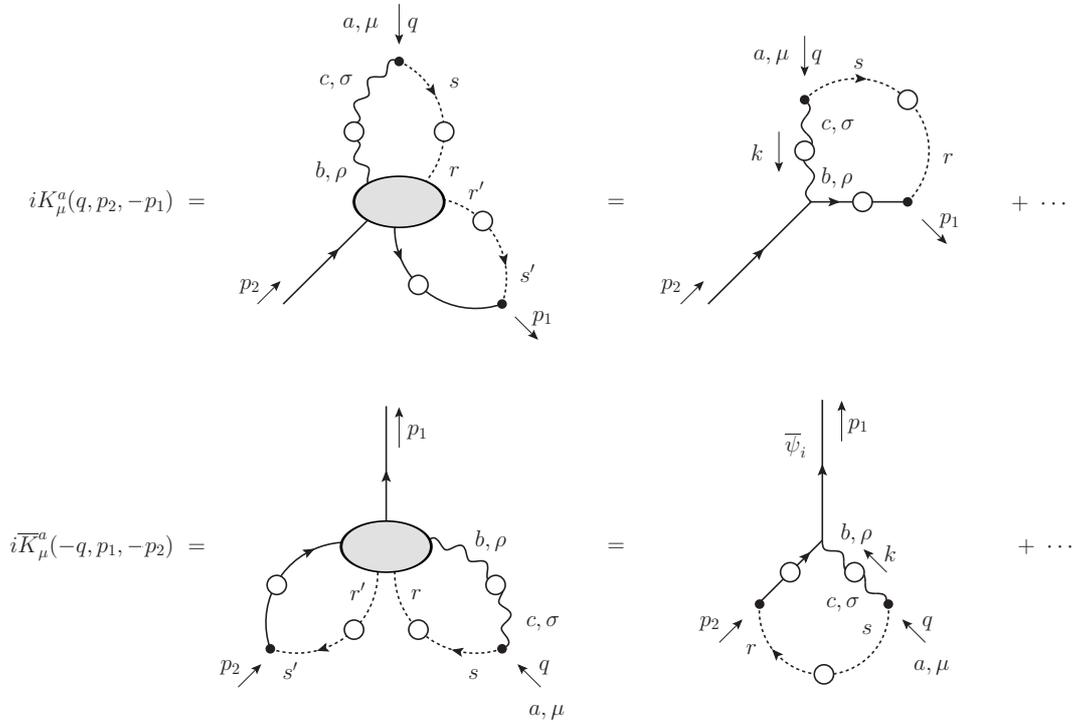}
\caption{\label{fig:K-functions}The auxiliary functions $K_\mu$ and $\overline{K}_\mu$ appearing in the BQI relating the conventional quark vertex $\Gamma_\mu$ with the PT-BFM vertex $\widehat{\Gamma}_\mu$. The diagrams on the right represent the one-loop dressed approximation of the two functions.}  
\end{center} 
\end{figure}

\begin{figure}[!t]
\begin{center} 
\includegraphics[scale=.8]{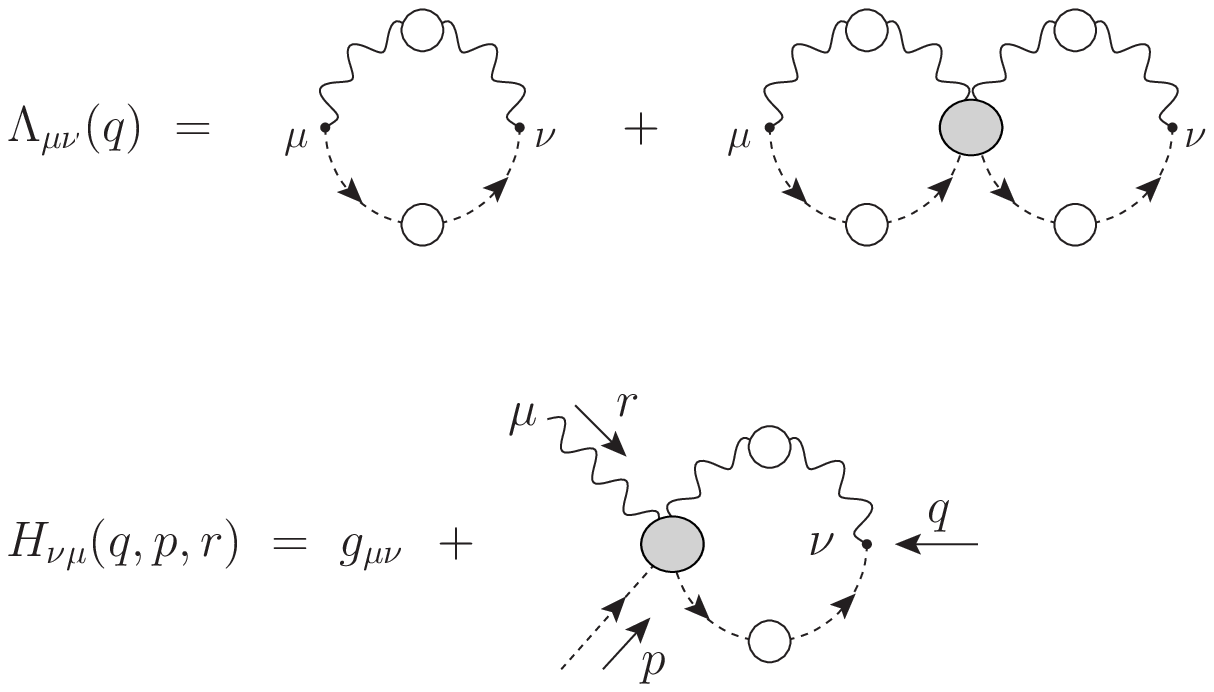}
\caption{\label{Lambda-H} The auxiliary function $\Lambda_{\mu\nu}$ appearing in the BQI relating the conventional quark vertex $\Gamma_\mu$ with the PT-BFM vertex $\widehat{\Gamma}_\mu$. The diagrammatic representation of the gluon-ghost scattering kernel $H_{\mu\nu}$. }
\end{center}
\end{figure}

In what follows we will use the one-loop dressed approximation, in which the propagators are fully dressed while vertices are retained at 
their tree-level (see \fig{fig:K-functions} again). This approximation yields the following expressions 
\begin{align}
K_\mu(q,p_2,-p_1)&=\frac i2g^2C_A\int_k\!S(k+p_2)\gamma^\nu P_{\mu\nu}(k)\Delta(k^2)D(k-q),\nonumber \\
\overline{K}_\mu(-q,p_1,-p_2)&=\frac i2g^2C_A\int_k\!\gamma^\nu S(p_1-k) P_{\mu\nu}(k)\Delta(k^2)D(k-q)\,,
\label{kkk}
\end{align}
where $C_\s A$ represents the Casimir eigenvalue of the adjoint representation [$C_\s A=N$ for SU(N)], 
$d=4-\epsilon$ is the space-time dimension, and we have introduced the integral measure 
 \mbox{$\int_k=\mu^\epsilon\int\!\diff{d}k/(2\pi)^d$}, 
with $\mu$ the 't Hooft mass.

The functions $G(q^2)$ and $L(q^2)$ appearing in  Eq.~(\ref{BQI})  
are the tensorial projections of the special two-point function 
\bea
\Lambda_{\mu\nu}(q)&=&-i\gA\int_k\!\Delta_\mu^\sigma(k)D(q-k)H_{\nu\sigma}(-q,q-k,k)\nonumber\\
&\equiv&g_{\mu\nu}G(q^2)+\frac{q_\mu q_\nu}{q^2}L(q^2),
\label{Lambda}
\eea
Finally, $H_{\mu\nu}$ is the ghost-gluon scattering kernel, and 
$\Delta_{\mu\nu}(q)$ is the gluon propagator defined in the Landau gauge 
as
\be
i\Delta_{\mu\nu}(q)=- i P_{\mu\nu}(q)\Delta(q^2), \qquad  P_{\mu\nu}(q)=g_{\mu\nu}- q_\mu q_\nu/q^2.
\label{Delta}
\ee

Interestingly enough, in the Landau gauge 
the form factors $G(q^2)$ and  $L(q^2)$ are related to the ghost dressing function $F(q^2)$ by the  all-order relation~\cite{Grassi:2004yq,Aguilar:2009pp}    
\be
F^{-1}(q^2) = 1 + G(q^2) + L(q^2).
\label{funrel}
\ee

The most general tensorial decomposition of the quark-gluon vertex contains
four longitudinal and eight transverse form factors. Using  
transverse and longitudinal (T+L) basis~\cite{Davydychev:2000rt,Kizilersu:1995iz}, we can write
\be
\Gamma^\mu(q,p_2,-p_1)=\sum_{i=1}^4\Gamma^{\s{L}}_i(q^2,p_2^2,p_1^2)L^\mu_i(q,p_2,-p_1)+\sum_{i=1}^8\Gamma^{\s{T}}_i(q^2,p_2^2,p_1^2)T^\mu_i(q,p_2,-p_1),
\label{bla}
\ee
where the longitudinal basis vectors read (remember that $t= p_1+p_2$)
\begin{align}
L^\mu_1&=\gamma^\mu;& L^\mu_2&=\tslash t^\mu;& L^\mu_3&=t^\mu;& L^\mu_4&=\widetilde{\sigma}^{\mu\nu}t_\nu;
\label{theLs}
\end{align}
while for the transverse basis vectors we have instead
\begin{align}
T^\mu_1&=p_2^\mu(p_1\cdot q)-p_1^\mu(p_2\cdot q);&
T^\mu_2&=T^\mu_1\tslash;\nonumber\\
T^\mu_3&=q^2\gamma^\mu-q^\mu\qslash;&
T^\mu_4&=T^\mu_1\widetilde{\sigma}_{\nu\lambda}p_1^\nu p_2^\lambda;\nonumber\\
T^\mu_5&=\widetilde{\sigma}^{\mu\nu}q_\nu;&
T^\mu_6&=\gamma^\mu(q\spr t)-t^\mu\qslash;\nonumber\\
T^\mu_7&=-\frac12(q\spr t)L^\mu_4-t^\mu\widetilde{\sigma}_{\nu\lambda}p_1^\nu p_2^\lambda;&
T^\mu_8&=\gamma^{\mu}\widetilde{\sigma}_{\nu\lambda}p_1^\nu p_2^\lambda+p_2^\mu\ps{1}-p_1^\mu\ps{2}.
\label{theTs}\end{align}

In addition to the usual WI~\noeq{WI} and STI~\noeq{STI}, specifying the \textit{divergence} of the quark-gluon vertex $\partial^\mu\Gamma_\mu$, there exists a set of less familiar identities, called transverse Ward identities (TWIs) \cite{Takahashi:1985yz,Kondo:1996xn,He:2000we,He:2006my,He:2007zza,Qin:2013mta}, which give information on the \textit{curl} of the vertex, $\partial_\mu\Gamma_\nu-\partial_\nu\Gamma_\mu$. 
 
In the case of an  Abelian gauge theory, one may consider a fermion coupling to a gauge boson 
through a vector vertex $\Gamma_\mu$ and an axial-vector vertex $\Gamma^\s{\rm A}_\mu$. In this case the TWIs read~\cite{Qin:2013mta}
\begin{align}
\label{TWI}
q_\mu \Gamma_\nu(q,p_2,-p_1) - q_\nu \Gamma_\mu(q,p_2,-p_1) &= i[S^{-1}(p_2)\widetilde{\sigma}_{\mu\nu} - \widetilde{\sigma}_{\mu\nu} S^{-1}(p_1)] +2im\Gamma_{\mu\nu}(q,p_2,-p_1)\nonumber \\
&+ t^\lambda \epsilon_{\lambda\mu\nu\rho}\Gamma^\rho_\s{\rm A}(q,p_2,-p_1) + A_{\mu\nu}^\s{\rm V}(q,p_2,-p_1),\nonumber \\
q_\mu \Gamma^\s{\rm A}_\nu(q,p_2,-p_1) - q_\nu \Gamma^\s{\rm A}_\mu(q,p_2,-p_1) &= i[S^{-1}(p_2)\widetilde{\sigma}_{\mu\nu}^5 - \widetilde{\sigma}_{\mu\nu}^5 S^{-1}(p_1)] \nonumber \\
&+ t^\lambda \epsilon_{\lambda\mu\nu\rho}\Gamma^\rho(q,p_2,-p_1) + V_{\mu\nu}^\s{\rm A}(q,p_2,-p_1),
\end{align}
where $\widetilde{\sigma}_{\mu\nu}^5=\gamma_5\widetilde{\sigma}_{\mu\nu}$, 
$\epsilon_{\lambda\mu\nu\rho}$ is the totally antisymmetric Levi-Civita tensor, 
while $\Gamma_{\mu\nu}$, $A_{\mu\nu}^\s{\rm V}$, and $V_{\mu\nu}^\s{\rm A}$ represent non-local tensor vertices that appear in this type of identities.

As was shown in Ref.~\cite{Qin:2013mta}, it is possible to disentangle  the vector and the axial-vector vertices appearing in Eq.~(\ref{TWI}). 
Specifically, for the vector vertex we obtain  
\be
[t^\mu\theta^i_\mu q_\rho - (q\spr t)\theta_\rho^i] \Gamma^\rho(q,p_2,-p_1) = P_i^{\mu\nu}\lbrace i[S^{-1}(p_2)\widetilde{\sigma}_{\mu\nu}^5 - \widetilde{\sigma}_{\mu\nu}^5 S^{-1}(p_1)] + V_{\mu\nu}^\s{\rm A}(q,p_2,-p_1)\rbrace, 
\label{transversei}
\ee
where the tensorial projectors are defined as 
\begin{align}
\label{proj}
P_i^{\mu\nu}& = \frac{1}{2}\epsilon^{\alpha\mu\nu\beta}\theta^i_\alpha q_\beta,\qquad i=1,2;& \theta^1_\alpha&=t_\alpha, \quad \theta^2_\alpha=\gamma_\alpha.
\end{align}

In what follows  we will use  Eq.~(\ref{transversei}) in conjunction 
with the WI~\noeq{WI} in order to  determine the complete set of form factors characterizing the vertex $\widehat\Gamma_\mu$. 
After that, we will apply the BQI given by Eq.~(\ref{BQI}) to obtain the final expression for $\Gamma_\mu$.

\section{Special kinematic configurations}

\bigskip
\indent

In this section, we will present the results for three special kinematic configurations: (i) the soft gluon limit, (ii) the symmetric limit and (iii) the zero quark momenta configuration. 
The general expressions, which are valid a generic configuration, can be found in Ref.~\cite{Aguilar:2014lha}.

\subsection{\label{soft} Soft-gluon limit }

\medskip
\indent

Let us start with the soft limit, obtained when we  take the limit of  $p_1\to p_2$  or similarly $q\to 0$. 
In this limit all the transverse tensor structures~\noeq{theTs} vanish identically. 
The vertex is therefore purely longitudinal, and after setting $p_1=p_2=p$,  the longitudinal tensorial structures  reduce ($p_1=p_2=p$) to
\begin{align}
L_1^\mu&=\gamma^\mu;&
L_2^\mu&=4\pslash p^\mu;&
L_3^\mu&=2p^\mu;&
L_4^\mu&=2\widetilde{\sigma}^{\mu\nu}p_\nu.
\end{align}

The form factors that accompany each one of the above tensorial structures are given by 
\begin{align}
F_0^{-1}\G{1}&=A\left(1-2p^2\K{4}{}\right)-2B\K{1}{},\nonumber\\
F_0^{-1}\G{2}&=2A'+2A\left(\K{3}{}+\K{4}{}\right)-2B\K{2}{},\nonumber \\
F_0^{-1}\G{3}&=-2B'+2A\left(\K{1}{}+p^2\K{2}{}\right)-2B\K{3}{},\nonumber \\
\G{4}&=0,
\label{softlimit-results-Mink}
\end{align}   
where $F^{-1}_0=F^{-1}(0)$, $A=A(p^2)$, $B=B(p^2)$, $\K{i}{}=\K{i}{}(p^2)$, and a prime denotes the derivative with respect to $p^2$. 
The  $\K{i}{}$ correspond to the decomposition of  the tensorial structure of Eq.~(\ref{kkk}) in the basis presented in Eq.~(\ref{bla}), and 
its detailed derivation is given in Ref.~\cite{Aguilar:2014lha}.

All ingredients that are necessary for computing the  $\K{i}{}=\K{i}{}(p^2)$
are renormalized at \mbox{$\mu=2.0$~GeV}. In particular, we use the  SU(3) gluon propagator, $\Delta(q)$, obtained  by the lattice simulation of Ref.~\cite{Bogolubsky:2009dc}, the solution of the SDE for ghost dressing function, $F(q)$, and the auxiliary functions $1+G(q)$ and $L(q)$. All these quantities where computed using  $\alpha(\mu)=0.45$.  In addition,  the behavior of the functions $A(p)$ and $B(p)$  were obtained 
by solving the quark gap equation  for a  current mass  $m_0=115 $ MeV.  

\begin{figure}[!h]
\centerline{\includegraphics[scale=.64]{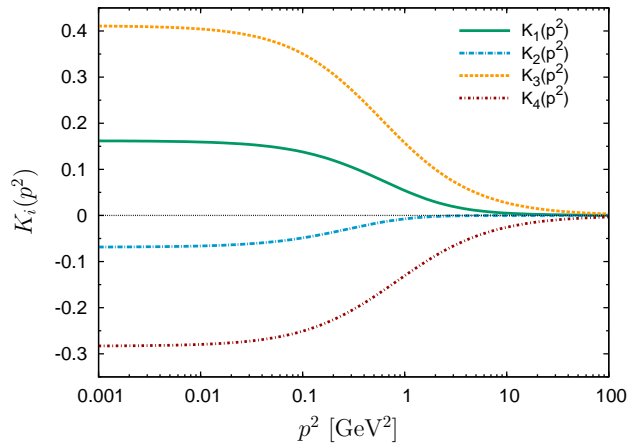}}
\caption{\label{fig:theKs_soft}(color online). The auxiliary functions $K_i$ evaluated in the soft gluon limit.} 
\end{figure}

In \fig{fig:theKs_soft} we plot the functions $K_i$ in the soft gluon limit. With the $K_i$  at hand,  the next step is to determine the vertex form factors of Eq.~(\ref{softlimit-results-Mink}).

\begin{figure}[!h]
\centerline{\includegraphics[scale=.995]{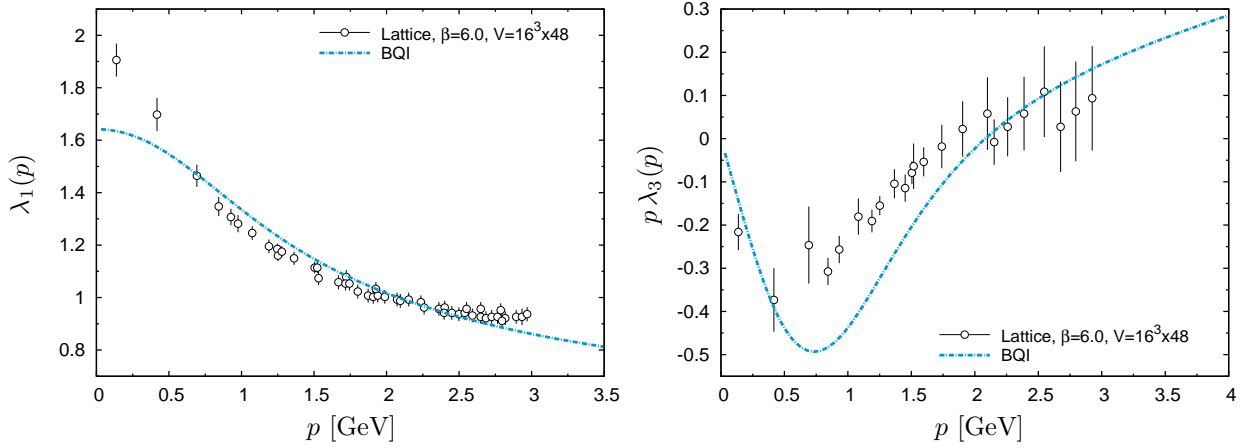}}
\caption{\label{fig:Gamma1-Gamma3_soft}(color online). 
The soft gluon form factors $\lambda_1$ (left) and $p\lambda_3$ (right). Lattice data in this and all the following plots are taken from~\cite{Skullerud:2003qu}.} 
\end{figure}

\begin{figure}[!h]
\centerline{\includegraphics[scale=.64]{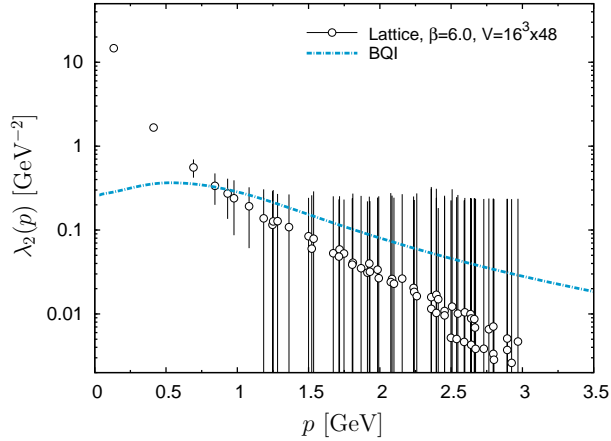}}
\caption{\label{fig:Gamma2_soft}(color online). The form factor $\lambda_2$ and the corresponding lattice data.} 
\end{figure}

Specifically, in~ Figs.~\ref{fig:Gamma1-Gamma3_soft} and \ref{fig:Gamma2_soft} we plot the form factors
\begin{equation}
\lambda_1(p)=\Ga{1}{\s{\rm E}}(p_\s{\rm E});\qquad 
\lambda_2(p)=\frac14\Ga{2}{\s{\rm E}}(p_\s{\rm E}); \qquad
\lambda_3(p)=-\frac12\Ga{3}{\s{\rm E}}(p_\s{\rm E}),
\end{equation}
and compare them with the lattice data of~\cite{Skullerud:2003qu}, obtaining rather reasonable agreement.


\subsection{Symmetric limit}

\medskip
\indent

The symmetric limit is defined taking $p_1\to-p_2$. In this limit,  
only one longitudinal basis tensor~\noeq{theLs} and two transverse tensors~\noeq{theTs} survive. Specifically we have
\begin{align}
L_1^\mu&=\gamma^\mu;&
T_3^\mu&=4\left(p^2\gamma^\mu-p^\mu\pslash\right);&
T_5^\mu&=-2\widetilde{\sigma}^{\mu\nu}p_\nu.
\end{align}

However, the lattice simulations can not determine separately  
the form factors that accompanies  the tensorial structures above described; only  combinations of the type 
\begin{align}
{\cal G}_{2p}(\Ga{1}{L}+p^2\Ga{3}{T})&=2p^2A'+A(1+2p^2\K{5}{T})-2B(\K{1}{L}+p^2\K{3}{T}),\nonumber \\
{\cal G}_{2p}\Ga{5}{T}&=2B'+2A\left(\K{1}{L}+p^2\K{3}{T}\right)-2B\K{5}{T}.
\label{symmff}
\end{align}
can be extracted. In the above equation 
the compact notation $A=A(p^2)$, $B=B(p^2)$ and  ${\cal G}_{2p}=1+G(4p^2)$ has been introduced.

\begin{figure}[!t]
\centerline{\includegraphics[scale=.65]{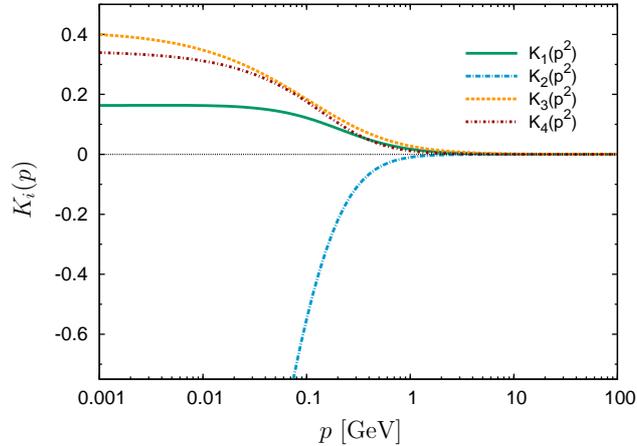}}
\caption{\label{fig:theKs_symm}(color online). The auxiliary functions $K_i$ evaluated in the symmetric gluon limit.} 
\end{figure}

Using the same ingredients described in the soft gluon case, we have 
computed the corresponding  $K_i$ for the symmetric configuration, which 
are presented in \fig{fig:theKs_symm}. In this limit, we  clearly see that we have a divergent $K_2$ and a finite $K_4$.

\begin{figure}[!h]
\centerline{\includegraphics[scale=.995]{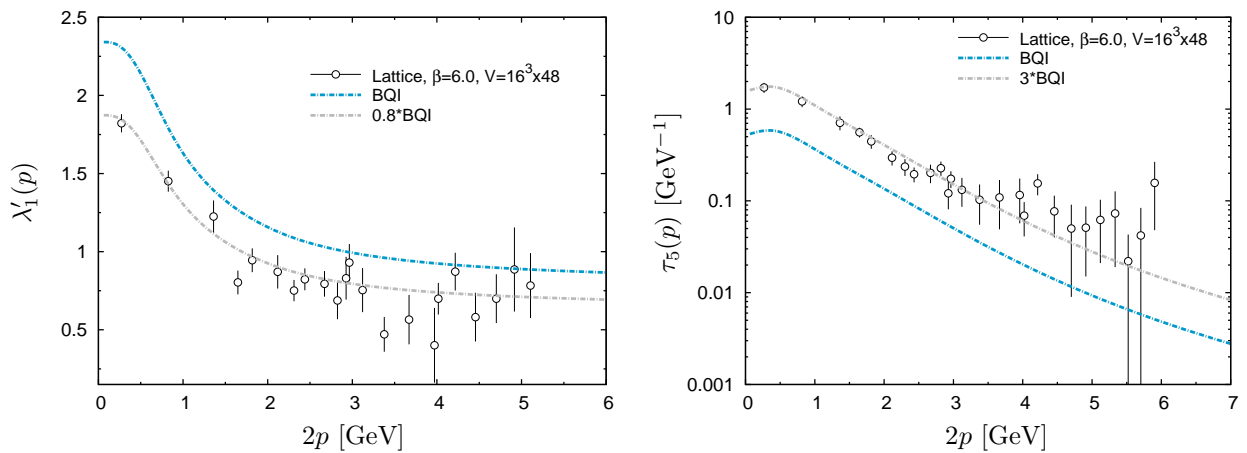}}
\caption{\label{fig:lambda1-tau5_symm}(color online). 
The symmetric limit form factors $\lambda'_1$ and $\tau_5$ compared with the corresponding lattice data.
The grey curves are obtained through simple rescaling of the blue ones.} 
\end{figure} 

The next step is to compare our numerical results for the Euclidean version of the form factors combinations of Eq.~(\ref{symmff}) with the  lattice data of Ref.~\cite{Skullerud:2003qu}.  This comparison is shown in \fig{fig:lambda1-tau5_symm}, where we have defined
\begin{equation}
\lambda'_1(p)=\Ga{1}{L\s{\rm{E}}}(p_\s{\rm{E}})-p^2_\s{\rm{E}}\Ga{3}{T\s{\rm{E}}}(p_\s{\rm{E}});\qquad 
\tau_5(p)=\frac12\Ga{5}{T\s{\rm{E}}}(p_\s{\rm{E}}) \,,
\label{prj-symm}
\end{equation}

Clearly,  we see that the overall shape of the both form factors are correctly reproduced; however, the overlap with the lattice data is only obtained 
if we rescale our results by (different) multiplicative factors, giving rise to the gray curves.

\begin{figure}[!h]
\centerline{\includegraphics[scale=.995]{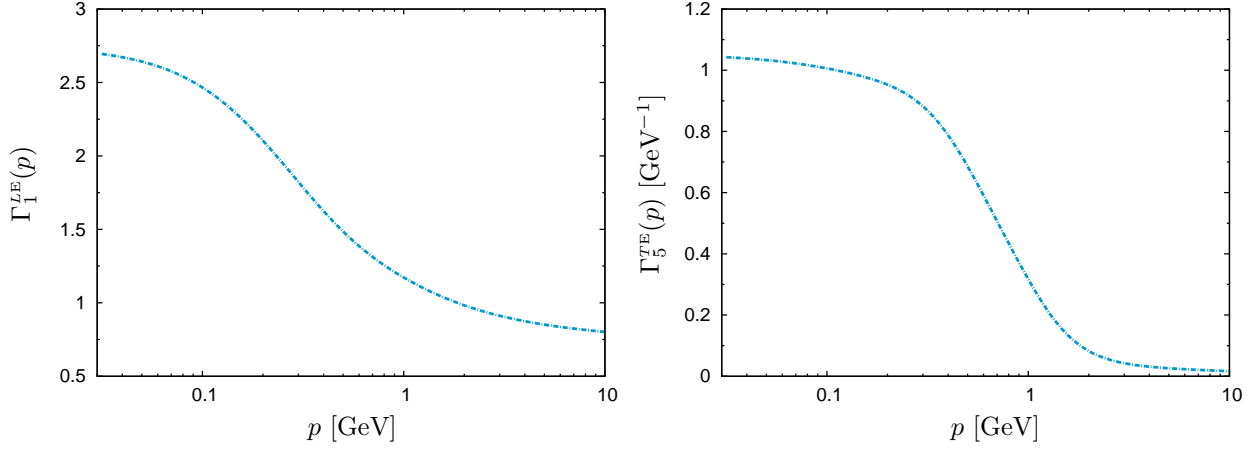}}
\caption{\label{fig:Gamma1L-Gamma5T_symm}(color online). 
The symmetric form factors $\Ga{1}{L}$ (left) and $\Ga{5}{T}$ (right).} 
\end{figure}

\begin{figure}[!h]
\centerline{\includegraphics[scale=.65]{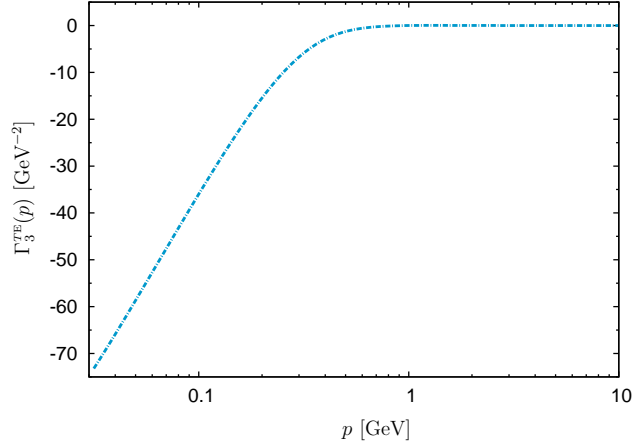}}
\caption{\label{fig:Gamma3T_symm}(color online). The divergent form factor $\Ga{3}{T}$ in the symmetric limit.} 
\end{figure}

In Figs.~\ref{fig:Gamma1L-Gamma5T_symm} and ~\ref{fig:Gamma3T_symm} we plot, for completeness the three non-zero form factors separately. Notice that 
both  $\Ga{1}{\s L}$ and $\Ga{5}{\s T}$ are finite, whereas  $\Ga{3}{\s T}$ is divergent. 
%
 
\subsection{Zero quark momentum}

\medskip
\indent

The next limit that we will present here  is the so-called zero quark configuration, where we set to zero the quark momentum $p_2$, which leads to  $q=p_1=p$. A crucial difference of this case compared with the previous one is that the $\K{i}{L,T}(p^2,0,p^2)$ and $\oK{i}{L,T}(p^2,p^2,0)$ do not coincide anymore, and need to be evaluated separately.  More specifically, 
the non-zero form factors are expressed as 
\begin{align} 
F^{-1} \Ga{1}{L} &= A(1 + p^2\K{3}{L}) - B\K{1}{L} - B_0\oK{1}{L}, \nonumber \\
F^{-1} \Ga{3}{L} &= -\frac{1}{p^2}(B-B_0) + A\K{1}{L} - B\K{3}{L} - B_0\oK{3}{L}, \nonumber \\
{\cal G}\Ga{3}{T} &=- A(\K{3}{L} + \K{5}{T}) - B\K{3}{T} - B_0\oK{3}{T} + \frac{1}{p^2}L_p F_p\left[A(1+p^2\K{3}{L}) - B\K{1}{L} - B_0\oK{1}{L}\right], \nonumber \\
{\cal G}\Ga{5}{T} &=  -\frac{1}{p^2}(B-B_0) - A(\K{1}{L}+p^2\K{3}{T}) - B\K{5}{T} - B_0\oK{5}{T},
\label{solBQIonep}
\end{align}
with the usual definitions $A=A(p^2)$,  $B=B(p^2)$, as well as $B_0=B(0)$.

\begin{figure}[!h]
\centerline{\includegraphics[scale=.995]{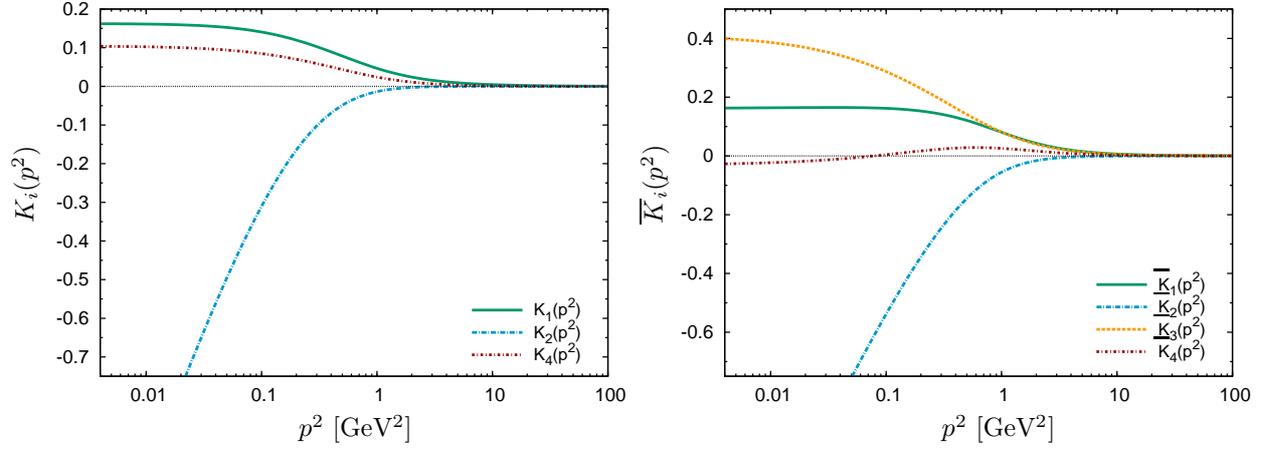}}
\caption{\label{fig:theKs-thebarKs_zero}(color online). 
The auxiliary functions $\K{i}{}$ (left) and $\oK{i}{}$ (right) evaluated in the zero quark momentum configuration.} 
\end{figure}

In Fig.~\ref{fig:theKs-thebarKs_zero} we show the auxiliary functions $\K{i}{}$ (left) and $\oK{i}{}$ (right) evaluated in the zero quark momentum configuration. Notice that all  $\K{i}{}$  and $\oK{i}{}$ are IR finite except for  $\K{2}{}$ and $\oK{2}{}$  which are divergent.

\begin{figure}[!t]
\mbox{}\hspace{-1.3cm}\includegraphics[scale=.995]{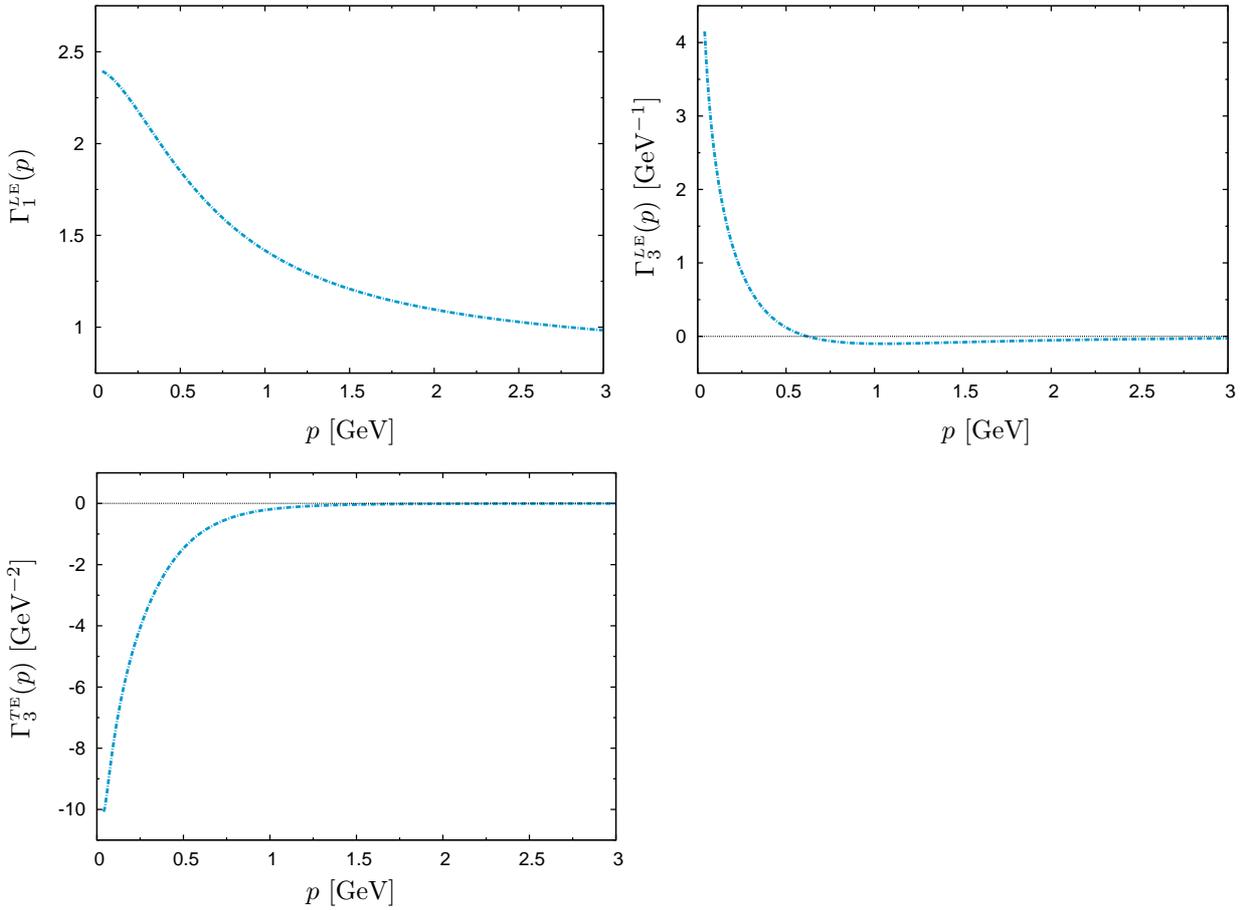}
\caption{\label{fig:allGamma_zero}(color online). 
The form factors $\Ga{1}{L\s{\rm E}}$, $\Ga{3}{L\s{\rm E}}$ and $\Ga{3}{T\s{\rm E}}$ evaluated in the zero quark momentum configuration.} 
\end{figure}

In \fig{fig:allGamma_zero} we show the Euclidean version of the 
numerical results for the  form factors appearing in Eq.~(\ref{solBQIonep}). In particular, we see the appearance of a  negative divergence  in the form factor $\Ga{3}{T\s{\rm E}}$.

Although there is no lattice results for this kinematic limit, 
the combination that could be measured on the lattice would be of the type
\begin{align}
{\cal G}(\Ga{1}{L}+p^2\Ga{3}{T})&=A(1-p^2\K{5}{T})-B(\K{1}{L}+p^2\K{3}{T})-B_0(\oK{1}{L}+p^2\oK{3}{T}),\nonumber \\
{\cal G}\Ga{5}{T}&=-\frac{1}{p^2}(B-B_0) - A(\K{1}{L}+p^2\K{3}{T}) - B\K{5}{T} - B_0\oK{5}{T}.
\label{ffzero}
\end{align}

\begin{figure}[!t]
\centerline{\includegraphics[scale=.995]{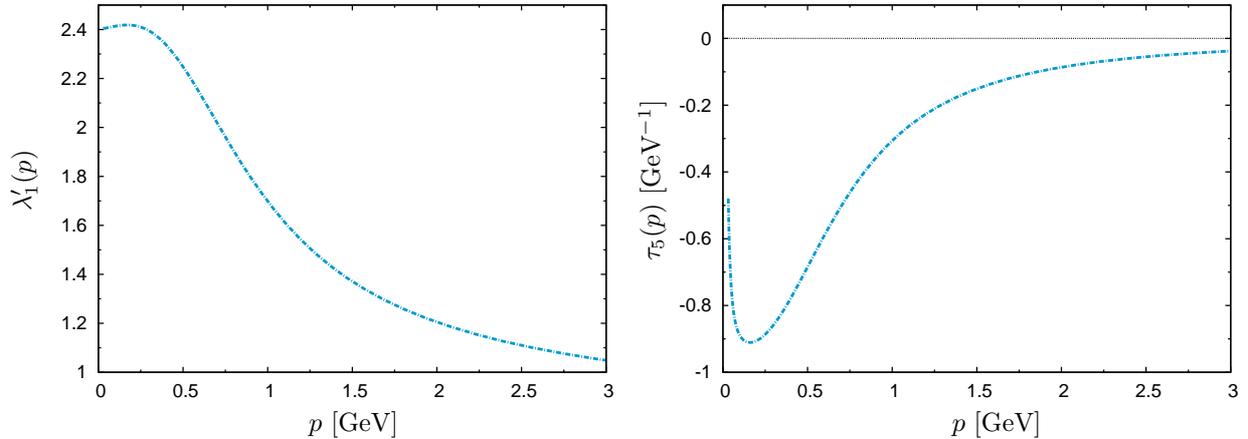}}
\caption{\label{fig:lambda1-tau5_zero}(color online). 
The form factors $\lambda'_1$ (left) and $\tau_5$ (right) in the quark zero momentum configuration.} 
\end{figure}

In this case, we present our  ``prediction'' for these combinations in Fig.~\ref{fig:lambda1-tau5_zero}. 

\section{Conclusions}

\bigskip
\indent

We have presented a new methodology for determining the longitudinal and transverse form factors of the nonperturbative quark-gluon vertex 
within the PT-BFM scheme. 
This scheme allows us to  take the full advantage of the rich amount of information originating 
from the fundamental underlying symmetries, which are encoded in a set of crucial identities such as WIs, STIs and BQIs. 

The key observation in this analysis is the connection between the two distinct
quark-gluon vertices, $\Gamma_{\mu}$ and $\widehat\Gamma_{\mu}$, appearing in the PT-BFM scheme. Using the WIs and the TWIs satisfied by $\widehat\Gamma_{\mu}$, we first determine the form factors that describes the behavior of this Abelian-like type of vertex. Then, with the help of the BQI 
that connects both vertices, we  obtain the final expression for the conventional $\Gamma_{\mu}$. 

We have shown that the BQI is expressed in terms of the auxiliary three-point functions $K_{\mu}$, which were calculated in the one-loop dressed 
approximation. Already, at this level of approximation, we have  obtained
nontrivial information for all form-factors. In addition, we have noticed that the  contributions originating from the $\K{i}{}$  and $\oK{i}{}$ are in general sizeable, and therefore 
the  $\K{i}{}$  and $\oK{i}{}$ contribute significantly in obtaining results 
similar to those found in lattice simulations. 

For the determination of the  $\K{i}{}$  and $\oK{i}{}$ we have used as external ingredient the full gluon propagator, $\Delta(q^2)$, 
obtained in lattice simulations. The remaining necessary ingredients, namely the ghost dressing function $F(q^2)$ and  the quark functions  $A(p^2)$ and $B(p^2)$,   were obtained  solving numerically  their corresponding  SDEs.

For the purpose of this talk we have applied our formalism 
to  three  particular kinematic limits known as (i) ``soft gluon", (ii)  ``quark symmetric" and (ii) ``zero quark momentum" configurations, which give rise to considerable technical simplifications, especially in the calculation of the  $\K{i}{}$  and $\oK{i}{}$.
Evidently, the numerical analysis presented here may be extended to arbitrary kinematic configurations, furnishing valuable information on such a fundamental quantity as the quark-gluon vertex, which constitutes a crucial ingredient for a variety of theoretical and phenomenological studies. 

\ack 

\bigskip
\indent

I would like to thank the organizers of the ``Discrete 2014'
for the pleasant conference and  for the hospitality. The work of  A.~C. Aguilar  is supported by the 
National Council for Scientific and Technological Development - CNPq
under the grant 306537/2012-5 and project 473260/2012-3,
and by S\~ao Paulo Research Foundation - FAPESP through the project 2012/15643-1.

\section*{References}

\bigskip


\providecommand{\newblock}{}

\end{document}